\documentclass[runningheads]{llncs}
\usepackage{graphicx}
\usepackage[ruled]{algorithm2e}
\usepackage{url}

\begin{document}
\title{Enabling Cross-chain Transactions: A Decentralized Cryptocurrency Exchange Protocol}

\titlerunning{Enabling Cross-chain Transactions: A Decentralized ...}

\author{Hangyu Tian$^{1}$ \and Kaiping Xue$^{1,*}$ \and Shaohua Li$^{2}$ \and Jie Xu$^{1}$ \and Jianqing Liu $^{3}$ \and Jun Zhao$^{4}$}

\authorrunning{Hangyu Tian, Kaiping Xue, et al.}

\institute{$^1$ University of Science and Technology of China, Hefei, Anhui 230027 China\\
$^2$ ETH Zurich, Zurich 8092, Switzerland\\
$^3$ University of Alabama in Huntsville, Huntsville, AL 35899 USA\\
$^4$ Nanyang Technological University, Singapore 639798\\
$*$ Corresponding Author, Email: kpxue@ustc.edu.cn}

\maketitle

\hyphenpenalty=5000
\tolerance=1000
\hyphenation{op-tical net-works semi-conduc-tor}

\begin{abstract}
Inspired by Bitcoin, many different kinds of cryptocurrencies based on blockchain technology have turned up on the market. Due to the special structure of the blockchain, it has been deemed impossible to directly trade between traditional currencies and cryptocurrencies or between different types of cryptocurrencies. Generally, trading between different currencies is conducted through a centralized third-party platform. However, it has the problem of a single point of failure, which is vulnerable to attacks and thus affects the security of the transactions. In this paper, we propose a distributed cryptocurrency trading scheme to solve the problem of centralized exchanges, which can achieve trading between different types of cryptocurrencies. Our scheme is implemented with smart contracts on the Ethereum blockchain and deployed on the Ethereum test network. We not only implement transactions between individual users, but also allow transactions between multiple users. The experimental result proves that the cost of our scheme is acceptable.

\end{abstract}

\section{INTRODUCTION}
In recent years, many blockchain-based cryptocurrencies have emerged, such as Bitcoin \cite{bitcoin}, Litecoin \cite{Litecoin} and Ethereum \cite{Ethereum}. To help users manage different kinds of cryptocurrencies, a centralized exchange \cite{P2PB2B,MXC,BKEX,Bilaxy,LBank} based on a trusted third party is commonly used. On the one hand, a centralized exchange provides users with convenience in fund management. On the other hand, a centralized exchange can also act as an intermediary to help users to trade between different types of cryptocurrencies.

Although centralized exchanges provide user convenience in trading, it also brings about some security risks. Once users keep their properties in a centralized exchange platform, it means that the exchange platform is the ``Archilles Heel'' of the system which could result in malicious use of users' properties and transaction information. For instance, Mt. Gox, used to be the world's largest bitcoin exchange, lost 850,000 bitcoins worth more than \$450 million in 2014 \cite{MtGox}. And over 43,000 bitcoins were stolen from the Bitcoinica trading platform in March 2012 \cite{Bitcoinica}. The loss of customers' information and assets also happened in many other exchange platforms \cite{lost}. Moreover, the centralized exchange platform is also subject to availability issue, and the work \cite{moore} shows that nearly half of the 80 exchange platforms established between 2010 and 2015 have closed. In light of these concerns, researchers have also done many works on the security of centralized exchange platform. Maxwell et al. \cite{max} proposed a protocol which allows users to verify whether their accounts are included in the liabilities of the exchange. The main idea is to use a binary Merkle hash tree and store each customer's balance in the leaf node. Each internal node contains the sum of the balance of its left and right child nodes. If a user wants to verify if his/her property is in the total liabilities of the exchange, the exchange only needs to provide the user with the portion of the Merkel tree that contains the path to the root. However, this scheme cannot protect users' privacy, as every query will reveal the total balance of the neighboring nodes. In order to solve the privacy protection problem in solvency proof, Provisions \cite{provisions} uses zero-knowledge proof, but it cannot guarantee security for future adversarial capabilities.

As a central institution, there will always be a single point of failure. The best way to solve this problem is to have a distributed cryptocurrency exchange scheme, which is also in line with the idea of decentralization of cryptocurrency. Smart contracts \cite{smart} are codes that can be deployed and executed on a blockchain. With smart contracts, we can implement many decentralized applications. Currently, Ethereum is the largest and most popular blockchain platform supporting the deployment of smart contracts. Users can send their smart contract codes to the Ethereum network through transactions, which can then be verified by miners and added to the blockchain. Any smart contract code saved in the blockchain can be invoked by users who meet certain conditions. In this paper, we consider using smart contracts based on Ethereum to implement a decentralized cross-cryptocurrency exchange scheme which can verify different types of cryptocurrency transactions sent by different users. We not only resolve transactions between individual users, but also consider the situation where multiple users trade with each other.

In this paper we make the following key contributions:
\begin{itemize}
\item We propose a decentralized cross-cryptocurrency exchange scheme based on smart contracts, which allows users to trade between different kinds of cryptocurrencies through different accounts. Users can initiate multiple transfers in a short period of time, and the proposed contract can collect multiple transfers from different users and complete these transactions at the same time.

\item We take an effective approach to validate different types of currency transactions using Ethereum smart contracts by selecting multiple verifiers and forming a committee within the group of untrusted users. Analysis shows that the committee can get the correct verification result for each transaction.

\item We implement and deploy our cross-cryptocurrency transaction scheme on the Ethereum test network and evaluate the running costs of each part of the contract on our local machine. Experimental results show that the local operation cost of our scheme is only related to the number of participants but unrelated to the number of transactions per user.
\end{itemize}

Our paper is organized as follows. In Section \ref{RELATED WORK}, we introduce a series of work related to our scheme. After we introduce our system and security model in Section \ref{model}, Section \ref{PROPOSAL} details our cross-cryptocurrency transaction scheme. Section \ref{SECURITY ANALYSIS} provides a security analysis of our scheme. Section \ref{IMPLEMENTATION} shows the detailed deployment of our scheme and the related performance analysis. Finally, we have a summary of our work in Section \ref{conclusion}.

\section{RELATED WORK}
\label{RELATED WORK}
In this section, we focus on the existing work on decentralized cryptocurrency trading schemes which are designed to solve the single point of failure problem existing in centralized exchanges.

The Metronome project \cite{Metronome} proposes a cryptocurrency called MTN that can be traded across different blockchains. While a user destroys the token on the source chain, he/she receives a proof of exit receipt that can be used on the target blockchain. However, Metronome can only be implemented in blockchains that support smart contracts, but cryptocurrencies that do not support smart contracts cannot be exchanged.

KyberNetwork \cite{Kybernetwork} is a highly liquid chain protocol that provides instant trading and redemption services for digital assets and cryptocurrencies which is currently deployed on Ethereum. KyberNetwork relies on the blockchain relay technology (such as BTCRelay \cite{BTCRelay}) to achieve cross-chain confirmation, so there are defects in the support of multiple currencies.

ERC-20 \cite{ERC-20} is a standard interface for tokens on the Ethereum blockchain. Some work \cite{0x,EtherDelta,IDEX,tBTC,wBTC} try to resolve transactions between ERC-20 tokens on Ethereum and bitcoin. The limitations with these solutions are that they only serve ERC-20 tokens on Ethereum. The Republic \cite{Republic} is a decentralized dark pool project between cryptocurrency pairs across different blockchains. Dark pool provides a hidden order book where financial assets and instruments are traded and matched by an engine built on a multi-party computation protocol. One disadvantage of Republic is that it only provides exchanges between bitcoin and tokens based on Ethereum. However, in our scheme, we use smart contracts on Ethereum to verify other types of cryptocurrencies through a verification committee selected from intermediary nodes, so we can support transactions between any different cryptocurrencies other than ether.

Both Cosmos \cite{cosmos} and Polkadot \cite{polkadot} are proposed to solve the interoperability of blockchains by using the Tendermint \cite{tendermint} consensus algorithm. Yet, the disadvantages of both two are that they only support the blockchain network that is compatible with them, and both need to have new tokens issued by the new network which increases the transaction burden.

XCLAIM \cite{xclaim} is a generic framework for cryptocurrency to achieve untrusted and efficient cross-chain switching. The disadvantage is that the currency on the issuing blockchain requires a contract that supports certain functions, so a limited-capacity scripting language such as Bitcoin does not support this operation. Tesseract \cite{Tesseract} describes how to mark existing cryptocurrencies with a trusted execution environment (TEE) to enable cross-chain transactions, while TEEs suffer from their own security concerns such as rollback \cite{ROTE} and side-channel attacks \cite{AsyncShock}.

Atomic Cross-Chain Exchange (ACCS) \cite{HerlihyAtomic,Cross-chain,MirazAtomic} based on hashed timelocks or signature locks \cite{Anonymous,A2l,Dlsag,EggerAtomic} enables secure cross-chain switching, but there are some limitations in practicality. For example, such an ACCS scheme is interactive, requiring all parties to be online, actively monitoring all relevant blockchains during execution, synchronizing clocks between blockchains, and relying on pre-established secure channels for communication. In addition, long waiting time is often incurred during transmission.

Due to these unsolved problems in the above schemes, we propose a cross-cryptocurrency transaction scheme based on smart contracts which not only supports transactions between existing cryptocurrencies but also enables fast payment of multiple transactions among users.

\section{System and Security Model}
\label{model}
\subsection{System Model}
The system model in this paper mainly contains the following components: payer, payee, intermediary, blockchain. In this section, we will briefly describe what these components represent and the overall architecture of our work.

The notations that will be used in the rest of our paper are summarized in Table \ref{tab1}.

\begin{table}
\centering
  \caption{Symbol definition.}
  \label{tab1}
  \setlength{\tabcolsep}{12.5mm}{
  \begin{tabular}{cc}
    \hline\hline
    Symbol   & Description \\
    \hline
    $\mathcal{A}$& The payer of a transaction \\
    $\mathcal{B}$& The payee of a transaction \\
    $\mathcal{C}_1$& The first intermediary of a transaction \\
    $\mathcal{C}_2$& The second intermediary of a transaction \\
    $coin_1$& Cryptocurrency owned by the payer \\
    $coin_2$& Cryptocurrency required by the payee \\
  \hline\hline
\end{tabular}}
\end{table}

\subsubsection{Components}
Our system mainly contains the following components:
\begin{itemize}
\smallskip
\item \textit{Payer.} The payer is a user who wants to transfer cryptocurrency to another user.

\smallskip
\item \textit{Payee.} The payee is a user who needs transfer-in, but the type of cryptocurrency he/she needs is not available on the payer's side.

\smallskip
\item \textit{Intermediary.} Some users act as intermediaries between payers and payees to realize transfers between different types of cryptocurrencies. Intermediaries need to connect payers and payees through smart contracts. The intermediaries need to join the validation committee to participate in the transaction validation process, which will be introduced in Section \ref{Trade validations}.

\smallskip
\item \textit{Blockchain.} Each execution of our scheme involves three blockchains. The two blockchains are the types of cryptocurrencies used respectively by the payer and payee. The third blockchain is the Ethereum blockchain, on which our smart contracts are designed and deployed.
\end{itemize}

We assume that each intermediary has the ability to verify different kinds of cryptocurrency transactions through wallet, blockchain browser and many other tools. Since the intermediary is required to verify transactions, it can be a full node or spv node \cite{bitcoin}. Provided that our smart contracts are deployed in Ethereum, we need all users participating in our scheme to have their own Ethereum accounts and a small amount of ethers which are used for calling contracts. Nevertheless, the amount of ethers held by the user cannot support his/her transactions. For users who only need to trade, they can join our network as a spv node.

\subsection{Security model}
Malicious users in our system may set up a large number of accounts to act as payers, payees, or intermediaries and send a large number of junk transactions to undermine our scheme. All malicious users may collude to gain extra benefits by defrauding the contract. Their possible malicious behaviors are shown below.
\begin{itemize}
\smallskip
\item Malicious payers or intermediaries may send spoofing messages to a contract without making transfers or make double spending.

\smallskip
\item Malicious payees or intermediaries can defraud the contract after receiving transfers to earn additional compensation fees.

\smallskip
\item Malicious nodes participating in the validation committee (description in the Section \ref{Trade validations}) may release wrong information to disrupt the consensus process.
\end{itemize}
Unlike malicious nodes, trusted nodes will comply with the terms of our protocol to enforce their behavior in order to achieve cross-cryptocurrency transactions or earn transaction fees through our scheme. Since the validation committee is elected by proof-of-work, we assume that in all nodes that want to join the validation committee, the combined hash power of the malicious nodes should be less than $1/4$ of the total hash power at any time. Otherwise, it will breed selfish-mining attacks \cite{selfish}.

\section{Cross-cryptocurrency transaction scheme}
\label{PROPOSAL}
In this section, we first explain how our scheme is applied to two users who trade with different cryptocurrencies. Then, we design a secure verification method to verify that the transaction is completed. Finally, we extend the scheme to the scenario where multiple users trade through different cryptocurrencies.

\subsection{Overview}
\begin{figure}
\centering
\includegraphics[width=0.65\linewidth]{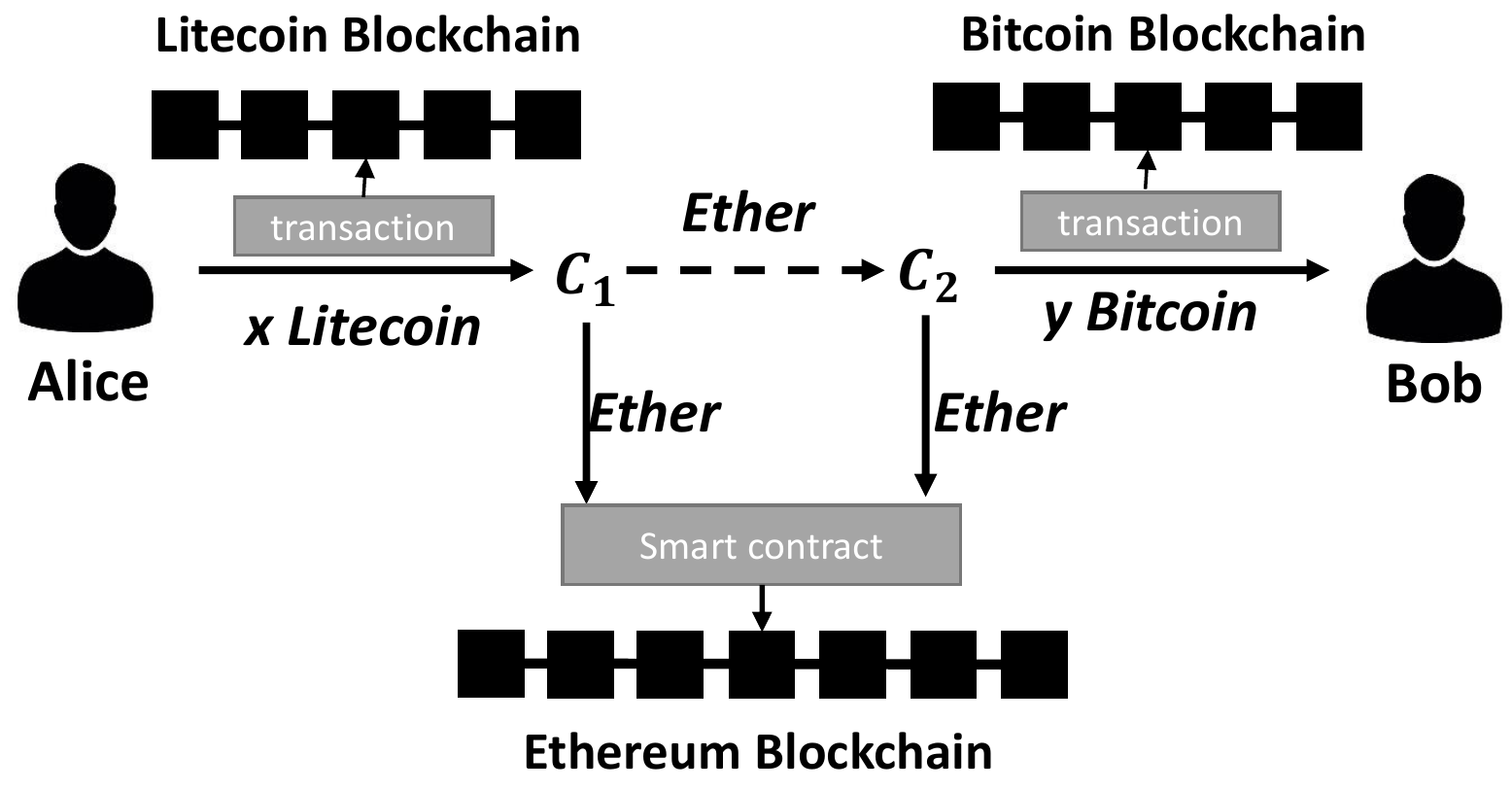}
\caption{An overview of the scheme.}
\label{fig1}
\end{figure}

In this section, we will give an overview of our scheme to show how we enable transactions across different kinds of cryptocurrencies. As we can see from Figure \ref{fig1}, to achieve a cross-cryptocurrency transaction through smart contracts, we need two intermediaries $\mathcal{C}_1$ and $\mathcal{C}_2$ who can be anyone in the network. Our requirement for intermediaries is that they need to support transactions through bitcoin and litecoin respectively. Firstly, $\mathcal{A}$ transfers $x$ litecoins to $\mathcal{C}_1$. After receiving the transfer, $\mathcal{C}_1$ transfers the equivalent ethers to $\mathcal{C}_2$, and then $\mathcal{C}_2$ transfers $y$ bitcoins to $\mathcal{B}$. This way, we use the transfer of ether as a bridge to achieve the transfer of different cryptocurrencies between two users. The incentive for the user to participate in our contract as an intermediary is that he/she can get transaction fees through this process.

It can be noticed that in addition to the transfer of ether in Ethereum, there is also transfer of bitcoin and litecoin. Although the transactions of these two currencies cannot be directly validated through the Ethereum smart contract, we find a solution by selecting a group of users as a validation committee to provide verification results. The contract will integrate the judgment results of the committee and draw the final conclusion. As concluded in work \cite{Sok}, cross-chain communication is impossible without a trusted third party, and the verification committee composed of distributed nodes plays such a role in our scheme.

In addition to the single-user transaction scenario mentioned above, we also consider the multi-user transaction scenario. In this case, there are multiple payers who want to transfer cryptocurrencies to one or many payees. Our contract could combine payers who need to transfer the same kind of cryptocurrency with a payee who needs a different kind of cryptocurrency. The details of our full proposal will be described in Section \ref{single} to \ref{multy}.

\subsection{Single-user Transaction Scheme}
\label{single}
The single-user transaction scheme represents that both parties involved in the transaction are individuals. This scheme also includes a user's currency exchange with himself/herself (e.g., Alice converts her bitcoin to ether). For clarity, we assume that the exchange rate between $coin_1$ and ether and the exchange rate between $coin_2$ and ether are both "1". In the real scenario, the problem of non-"1" exchange rate only needs to multiply the rate with the corresponding magnification. The main procedures of the single-user transaction scheme are shown in Algorithm \ref{single1}.

  \begin{algorithm}[htb]
  \caption{Framework of one-to-one trading scheme.}
  \label{single1}
  \KwIn
    {
      The account addresses of both parties and the intermediary; Transaction amount $x$; Exchange rate;}
    \If {Both intermediaries $\mathcal{C}_1$ and $\mathcal{C}_2$ have deposited Ether}{
         Payer $\mathcal{A}$ transfers $x coin_1$ to intermediary $\mathcal{C}_1$\;
        \eIf {Transfer confirmed successful}{
             Intermediary $\mathcal{C}_2$ transfers $x coin_2$ to payee $\mathcal{B}$\;
            \eIf {Transfer confirmed successful}{
                 Return the deposit of intermediary $\mathcal{C}_2$ to himself\;}
            {
                 Transfer the deposit of intermediary $\mathcal{C}_2$ to $\mathcal{B}$;
            }
             Transfer the deposit of intermediary $\mathcal{C}_1$ to $\mathcal{C}_2$\;
        }{
             The deposit of intermediary $\mathcal{C}_1$ and $\mathcal{C}_2$ are returned\;
        }
    }
\end{algorithm}

Before the transaction begins, the intermediaries issue their own supported cryptocurrency types and corresponding exchange rates to the contract. The parties of the transaction including $\mathcal{A}$ and $\mathcal{B}$ select the appropriate intermediaries $\mathcal{C}_1$ and $\mathcal{C}_2$ and publish information about their transaction to the contract, such as cryptocurrency types, transaction amount and account addresses. Then, the intermediaries $\mathcal{C}_1$ and $\mathcal{C}_2$ receive information about the transaction of both sides and notify contract about the users they select. Then, they need to send a certain amount of deposit to the contract. After receiving the message that payer $\mathcal{A}$'s transfer to $\mathcal{C}_1$ has succeeded, the intermediary $\mathcal{C}_2$ transfers $x\ coin_2$ to payee $\mathcal{B}$. Otherwise, the contract returns the deposit of $\mathcal{C}_1$ and $\mathcal{C}_2$ and terminates the transaction. If the final transfer transaction is successful, the deposit of $\mathcal{C}_2$ will be returned to himself. Otherwise, the deposit saved in the contract by the intermediary $\mathcal{C}_2$ will be sent to $\mathcal{B}$. In the end, the intermediary $\mathcal{C}_2$ will receive the deposit of $x\ ethers$ from $\mathcal{C}_1$.

\subsection{Trade Validation}
\label{Trade validations}
The transfer between the intermediary $\mathcal{C}_1$ and $\mathcal{A}$ and the transfer between the intermediary $\mathcal{C}_2$ and $\mathcal{B}$ need to be confirmed. And the validation result needs to be sent to the contract for the next step. We cannot simply ask any user to participate in the transaction validation because malicious users may masquerade as both sides of the transfer. For example, it is possible for the payer to send a fraudulent message to the contract about a successful transfer without any actual transfer. Similarly, the payee may send a message that the transfer failed after receiving the currency. In our scheme, we design a reliable method to remedy this problem by decentralizing the authority of transaction validation.

We consider that there are a large number of intermediaries in the system, but only two of them are needed for each transaction progress. Therefore, the rest of intermediary nodes could be used in our scheme as validators to verify the transaction. The intermediary earns fees by participating in the validation process. We also require each node to participate in a successful validation process at least once before becoming an intermediary. We assume that each intermediary has the ability to verify the transfer of different types of cryptocurrencies through wallet, block explorer and many other tools. In general, different types of cryptocurrencies have their own confirmation policies. For example, in Bitcoin, recipients assume their transactions are secure with 6 blocks attached.

\begin{figure}
\centering
\includegraphics[width=0.85\linewidth]{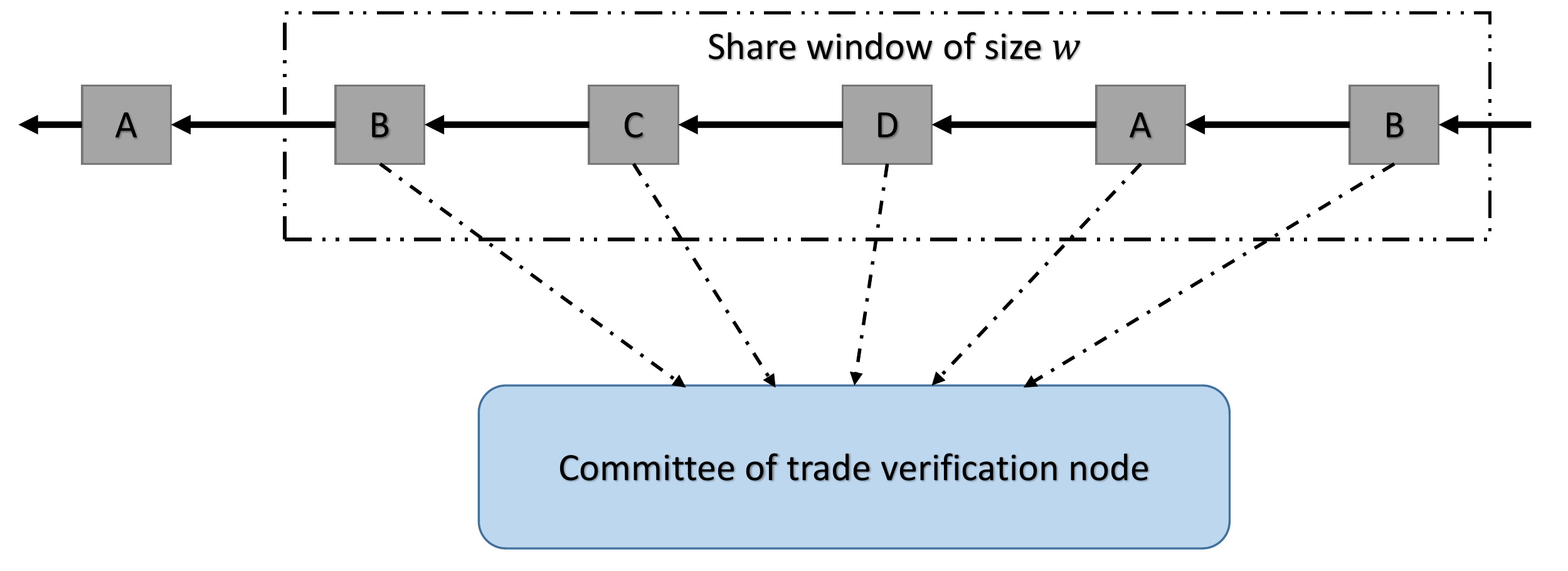}
\caption{Formation of a transaction validation committee.}
\label{fig_committee}
\end{figure}

To prevent malicious users from registering multiple intermediary accounts to attack the validation process, we use the proof-of-work(PoW) algorithm to determine the set of intermediaries participating in the validation committee. Intermediaries who resolve the proof-of-work puzzle are elected to the validation committee. We set the difficulty of the proof-of-work puzzle such that each of the two intermediaries participating in committee takes about $10$ minutes to solve. This way, all transactions generated within $10$ minutes will be validated by the intermediary in this committee. Before joining the validation committee, each intermediary who successfully solves the puzzle of PoW is required to pay some ethers as a security deposit. If an intermediary node is always honest, it will not only get back its deposit but also get rewards from the penalty of dishonest nodes and the transaction fee from the guarantee of the transaction. Thus, being an intermediary node in our system is monetarily incentivized. The formation of the trade validation committee is shown in Figure \ref{fig_committee}.

After each transaction, every node in the validation committee needs to verify the transaction and sends its own validation results to the smart contract. The smart contract records the address of these nodes and counts the validation results. Then, the validation committee will classify nodes that give conflicting validation results. We use the majority rule to determine the final validation result. To this end, the nodes whose results conflict with the final result will be considered dishonest and thus are excluded from the validation committee. Their deposit will be equally distributed to the honest nodes. The correct result will serve as the basis for the next step of smart contracts. We have a quantitative limit of $w$ for the size of the validation committee. The reason for this is that by enforcing the committee size, all intermediaries' deposit are kept in the contract till any dishonest node is excluded from the validation committee. Similar to a timing policy, conforming a committee size allows the deposit of dishonest nodes to be transferred to the honest nodes once node exclusion is triggered.

\subsection{Multi-user Transaction}\label{multy}
In many cases, there may be multiple users trading simultaneously in a short period of time. Our smart contract is designed to support multiple users to participate in cryptocurrency transfer. In our scheme, multiple users can individually select their own trading partners. The contract will combine these transaction information and aggregate the amount of cryptocurrency transferred to the same user to improve transaction efficiency.

 \begin{algorithm}[htb]
  \caption{ Framework of multi-user trading scheme.}
  \label{multi-user}
    \KwIn{
      The account address of payers: $\mathcal{A}_1,\mathcal{A}_2,\mathcal{A}_3......\mathcal{A}_n$; payees: $\mathcal{B}_1,\mathcal{B}_2,\mathcal{B}_3......\mathcal{B}_n$ and the intermediary:$\mathcal{C}_1$, $\mathcal{C}_2$; Total transaction amount:$t$; Amount to be sent by payers $\mathcal{A}_i$:$amount_A[i]$; Amount to be received by payees $\mathcal{B}_i$:$amount_B[i]$;}
      Set each element in the array $amount_B$ to zero\;
    \If {Both intermediary $\mathcal{C}_1$ and $\mathcal{C}_2$ have deposited Ether}{
         All the payers $\mathcal{A}_1,\mathcal{A}_2,\mathcal{A}_3......\mathcal{A}_n$ transfer $coin1$ to intermediary $\mathcal{C}_1$\;
        \For{$i=1$ to $n$}{
            \eIf {Sender $\mathcal{A}_i$ transfers successfully}{
                 Record $\mathcal{A}_i$\;
                 Add the value $amount_A[i]$ to the array element $amount_B[j]$ corresponding to the payee $\mathcal{B}_j$ of $\mathcal{A}_i$\;}
            {
                 $t=t-amount_A[i]$; Transfer the deposit equivalent to $amount_A[i]$ to $\mathcal{C}_1$ and $\mathcal{C}_2$\;

            }
        }
             Intermediary $\mathcal{C}_2$ transfers $coin2$ to the payees $\mathcal{B}_1,\mathcal{B}_2......\mathcal{B}_n$ separately\;
            \For{$i=1$ to $n$}{
                \eIf {The transfer to $\mathcal{B}_i$ is successful}{
                     Record $\mathcal{B}_i$\;
                     The deposit of $\mathcal{C}_2$ equivalent to $amount_B[i]$ will be returned to $\mathcal{C}_2$\;}
                {
                     Transfer the deposit of $\mathcal{C}_2$ equivalent to $amount_B[i]$ to $\mathcal{B}_i$\;
                }
            }
             Transfer the remaining deposit of quantity $t$ of intermediary $\mathcal{C}_1$ to $\mathcal{C}_2$\;
    }
\end{algorithm}

The framework of multi-user trading scheme is shown in Algorithm \ref{multi-user}. This scheme is mainly used in the scenario where a group of users trade with each other. Smart contract first collects information from all payers $\mathcal{A}_1,\mathcal{A}_2,\mathcal{A}_3......\mathcal{A}_n$  and works out the sum of the transaction amount. After the intermediaries $\mathcal{C}_1$ and $\mathcal{C}_2$ receive this information, they are required to pick the appropriate users and submit the equivalent of ethers as deposit. Then, all the payers needs to pay enough $coin_1$ to $\mathcal{C}_1$. Members of the validation committee need to verify these transfers through the methods described in Section \ref{Trade validations}. The transfer information will also be recorded in the contract. After the intermediary $\mathcal{C}_1$ receives the transfer from the payers, the contract calculates the amount of successful transactions, modifies the amount received by $\mathcal{C}_1$, and then sends the message to $\mathcal{C}_2$. The intermediary $\mathcal{C}_2$ finally transfers $coin_2$ to each of the payees $\mathcal{B}_1,\mathcal{B}_2,\mathcal{B}_3......\mathcal{B}_n$ based on the received message. Finally, if the transfer to payee $\mathcal{B}_i$ is judged to fail, the contract will send the same amount of the intermediary $\mathcal{C}_2$'s deposit to $\mathcal{B}_i$. At the same time, the contract will also send the remaining deposit of intermediary $\mathcal{C}_1$ to $\mathcal{C}_2$.

\subsection{Contract Implementation}
\label{function}
In our scheme, there are two smart contracts written in Ethereum's Solidity language. The first one is called the intermediary contract. As the name suggests, this smart contract primarily controls the behavior of the intermediary. The second contract is called the transaction contract, by which the payer and payee of the transaction complete the cross-cryptocurrency transaction with the participation of intermediaries.

The main functions of our contracts are listed as follows. We denote intermediary contract as IC and transaction contract as TC.

\begin{table}
  \caption{Smart contract function.}
  \label{tab2}
  \setlength{\tabcolsep}{3mm}{
  \begin{tabular}{cc}
    \hline\hline
    Contract steps   & Function \\
    \hline
    \textit{IC.Register}& Intermediary provides registration information \\
    \textit{IC.Update}& Intermediary update information \\
    \textit{IC.Verify\_PoW}& Verify wether the user is eligible to join the verification group \\
    \textit{TC.Prepare}& Information preparation for trading users\\
    \textit{TC.Deposit}& User submits deposit \\
    \textit{TC.validation}& User's transaction verification \\
  \hline\hline
\end{tabular}}
\end{table}

\section{SECURITY ANALYSIS}\label{SECURITY ANALYSIS}
In this section, we provide a security analysis of our scheme, and discuss how our scheme can mitigate or eliminate some well-known attacks against blockchain.
\subsection{Security of Validation Committee}
During the verification phase of the protocol, there might be malicious nodes in the verification committee who may provide incorrect verification results to disrupt the execution of our protocol. In our scheme, the validation committee is formed through proof-of-work. We assume that in all nodes that want to join the validation committee, the combined hash power of the malicious nodes should be less than $1/4$ of the total hash power at any time. Otherwise, there will be selfish-mining attacks \cite{selfish}. Suppose a total of $w$ nodes are selected to join the validation committee. If we require that the count of the final validation result from the majority rule exceeds $a$, it means that the percentage of malicious nodes in the validation committee is enforced to be less than $t=\frac{w-a}{w}$. We assume that $X$ is a random variable that represents the number of times we pick a Byzantine node in the validation committee. In this way, the maximum number of malicious nodes $c$ is equal to $\left\lfloor{w}\times{t}\right\rfloor$. We can use the cumulative binomial distribution to calculate the probability $P$ when the proportion of malicious node is less than $t$ in the committee of $w$ nodes.
$$P[X\le c]=\sum_{k=0}^c\Big(_k^w\Big)p^k(1-p)^{w-k}.\eqno(1)$$

\begin{table}
\centering
  \caption{The relationship between the security level of the validation set and its membership and the proportion of Byzantine nodes.}
  \label{tab3}
 \setlength{\tabcolsep}{8mm}{ \begin{tabular}{cccccc}
    \hline\hline
    $t/w$   & 10    & 20    & 50    & 100   \\
    \hline
    0.7   & 0.9997 & 1.0000 & 1.0000 & 1.0000 \\
    0.6   & 0.9965 & 1.0000 & 1.0000 & 1.0000 \\
    0.5   & 0.9957 & 0.9992 & 1.0000 & 1.0000 \\
    0.4  & 0.9219 & 0.9784 & 0.9937 & 0.9997 \\
    0.3  & 0.7759 & 0.8034 & 0.8369 & 0.8962 \\
  \hline\hline
\end{tabular}}
\end{table}

It can be seen from Table \ref{tab3} that when we take 10 validation committee members, the probability that the number of malicious nodes exceeds 1/2 of the total number is 0.99. This means that if the smart contract receives a consistent judgment from more than half of the validation committee members, there is a probability of 0.99 that the result is correct. From the table, we know that the higher the number of members in the validation committee, the higher the probability that the contract will get the correct judgment result after collecting enough validations. On the other hand, the more the contract receives the same validation result, the accuracy of the judgment results will also increase. If the number of validators in the validation committee reaches 100, the probability of receiving an accurate result is 0.99 after the contract receives a consistent result from more than 4/10 of the members. That is because the number of malicious nodes has a probability of 0.99 to be less than 10. This also means that for our scheme, under the premise of not reducing the security level, the number of the same judgments that the contract needs to collect can be reduced by increasing the number of validation committee members.

\subsection{Double-Spend Attack}
In our scheme, there is a situation where the payer entrusts two different intermediaries to make the same payment to different users at the same time. Or in the process of entrusting the intermediary, the payer may make the same payment on other occasions. In order to prevent double-spending attack \cite{Double-spending}, our scheme requires the verifier to comply with the validation method of the cryptocurrency itself. For example, if the payer pays in bitcoin, the verifier must wait for more than 6 blocks after the transaction before sending its judgment to the contract. Our scheme does not make any changes to the cryptocurrency involved to ensure the security of cryptocurrency transactions.

\subsection{Sybil Attack}
Any user can send a transaction request, even if the user does not make subsequent transfers after sending the application to the contract. In this case, only a small amount of the Ethereum transaction fees will be lost. During this process, malicious users could send a large amount of small transaction requests and may even refuse to initiate a transaction after requests are sent. However, in our scheme, the intermediary is allowed to make a selection after receiving the user's transaction request. That is to say the intermediary can confirm that the account has sufficient balance through the user's account address. The intermediary will remove the user request if the transfer amount is too small, thereby ensuring that the user is not a Sybil node \cite{sybil-attack}.

\subsection{DoS Attack}
A malicious node may forge a large number of accounts trying to join the validation committee or pretend to be an intermediary in order to disrupt the normal protocol process through DoS attacks. However, our scheme will not be affected by these malicious behaviors since we have two preventive measures. First, each member of the validation committee should pay some ethers as security deposit. Second, each node needs to submit a solution to the PoW puzzle in order to be recognized as a member of the validation committee. Therefore, a malicious node cannot easily forge its identity to join the validation committee without paying any cost. Likewise, a malicious node cannot act as an intermediary because the intermediary node needs to have a record of having joined a validation committee.

\section{PERFORMANCE ANALYSIS}
\label{IMPLEMENTATION}
To evaluate the deployment cost of our scheme, we deploy our contract on the Ethereum's official test network \cite{Ropsten}. Gas is used to measure the cost of each step in the smart contract code. Gas and financial costs of each of functions described in Section \ref{function} are outlined in Table \ref{tab4}. We have approximated the cost in USD (\$) using the conversion rate of 1 ether = \$130 and the gas price of 0.000000003 ethers which are the real cost in April 2020. The code for the intermediaries to calculate the PoW puzzle is not included in the contract because users can execute this part on their local machine and send the results to the contract.

\begin{table}
\centering
  \caption{The gas value of each operation.}
  \label{tab4}
  \setlength{\tabcolsep}{8mm}{\begin{tabular}{ccc}
    \hline\hline
    Transaction & Gas & USD \\
    \hline
    \textit{Deploy}   & 3690283 & 1.43\\
    \textit{IC.Register}   & 181591 & 0.07\\
    \textit{IC.Update}   & 80995 & 0.03\\
    \textit{IC.Verify\_PoW}  & 191737 & 0.07\\
    \textit{TC.Prepare}  & 398525 & 0.15\\
    \textit{TC.Deposit}  & 36452 & 0.01\\
    \textit{TC.Validation}  & 163780 & 0.06\\
    \hline\hline
\end{tabular}}
\end{table}
For users who need to make cross-currency transactions, only the function of \textit{TC.Prepare} needs to be invoked by the user. Even for the multi-party situation, each user only needs to call this function once, which costs \$0.15. For intermediate users, there are \textit{IC.Register} and \textit{IC.Update} functions to provide their own information, which cost only \$0.07 and \$0.03 each. In a transaction, the operation of issuing deposit for each intermediary will cost \$0.01 in average. The intermediary who wants to join the validation group needs to perform the validation operation of \textit{IC.Verify\_PoW} costing \$0.07. The cost of validation is borne by every member of validation group. Depending on the invocation overhead of our contract, users can choose whether to use our service within acceptable limits.

\subsection{Timing Analysis}
We also measure the runtime of our contract. All measurements were performed on a desktop running Windows 10 equipped with 6 cores, 3.0 GHz Intel Core i5 and 8 GB DDR4 RAM. In this section, we will present the runtime results for our scheme, which shows that our contract can achieve the expected functionality within an acceptable time overhead.

The implementation of our scheme is divided into two parts: off-chain part and on-chain part. The calculation of PoW puzzle performed by users to join the verification group is performed off-chain on the user's local machine. The smart contract only needs to receive and verify the solution of the PoW puzzle through \textit{IC.Verify\_PoW} function and adjust the difficulty value of PoW according to the solution time. In this way, the time interval for solving the PoW problem can be kept at around ten minutes. For the on-chain part, we only consider the operating time of the smart contract in the scheme. The scheme involves the confirmation time of other types of cryptocurrencies on its blockchain, which is not our consideration.

\begin{figure}
\centering
\includegraphics[width=0.6\linewidth]{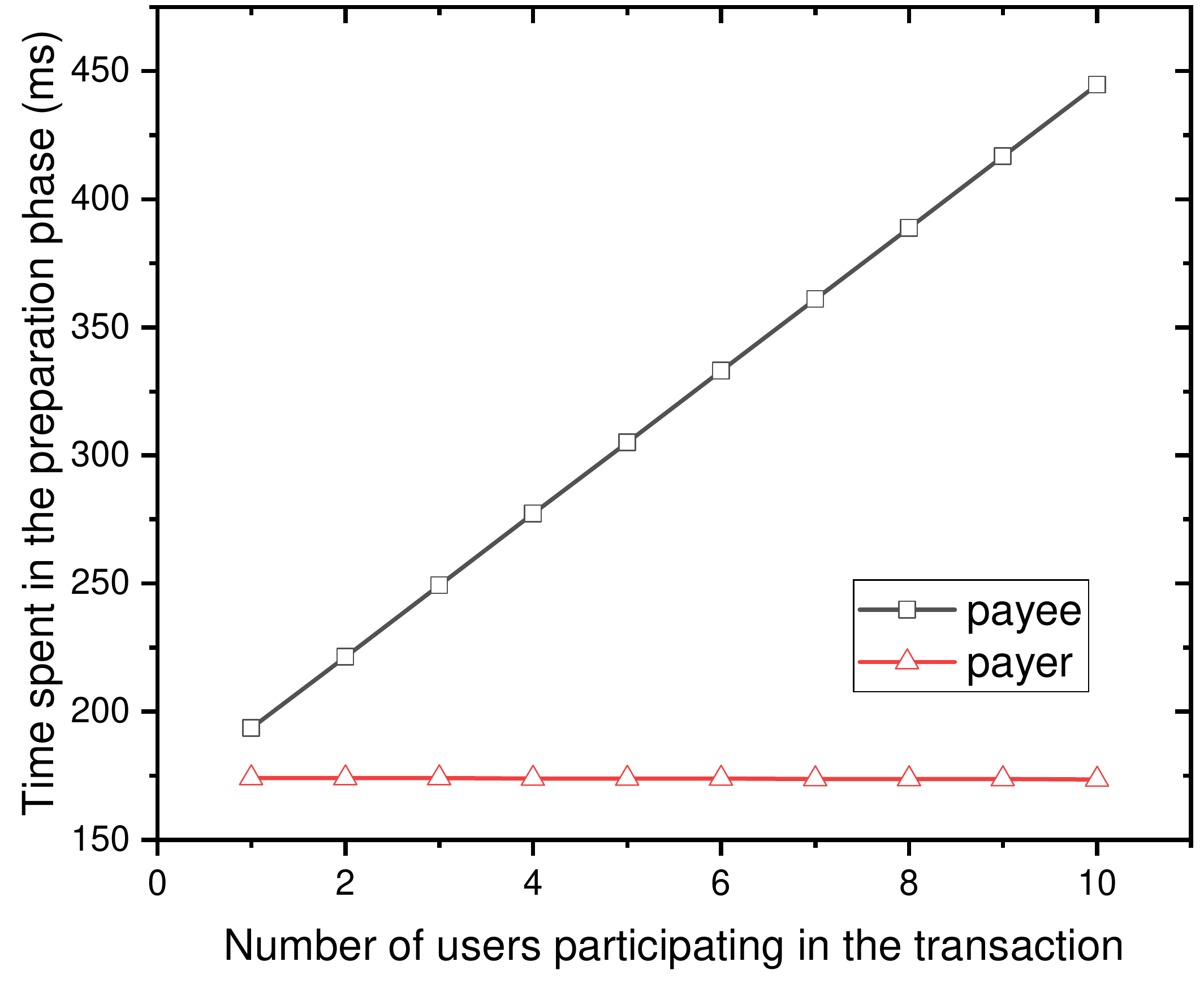}
\caption{Time spent on the preparatory phase when the number of payees increases.}
\label{figpayee}
\end{figure}

Due to the limitation of gas and block size, we set the number of payers and payees involved in the transaction to be less than ten in a round of protocol process. When we change the number of users participating in the transaction, the execution time of the preparation phase of the contract will be mainly affected. We record the largest execution time of the users who independently call the contract through \textit{TC.Prepare} function and show the results in Figure \ref{figpayee}. As we can see, the runtime of our contract increases as the number of payees increases. This is because for a single payer, the contract needs to record and process the information of each payee of the user. Thus, the larger the number of all payees, the more information will be sent to the contract, which leads to a longer overall runtime of the contract. It can be also seen that since each of the payers is a different node and their operations are parallel. Increasing the number of payers does not significantly increase the contract runtime. Thus, the preparation time of the contract is only related to the number of transaction users of each payer. After collecting ten users' requests or exceeding this waiting time, the contract will proceed to the next step.

\begin{figure*}
\centering
\begin{minipage}[t]{0.495\linewidth}
    \includegraphics[width=1.0\linewidth]{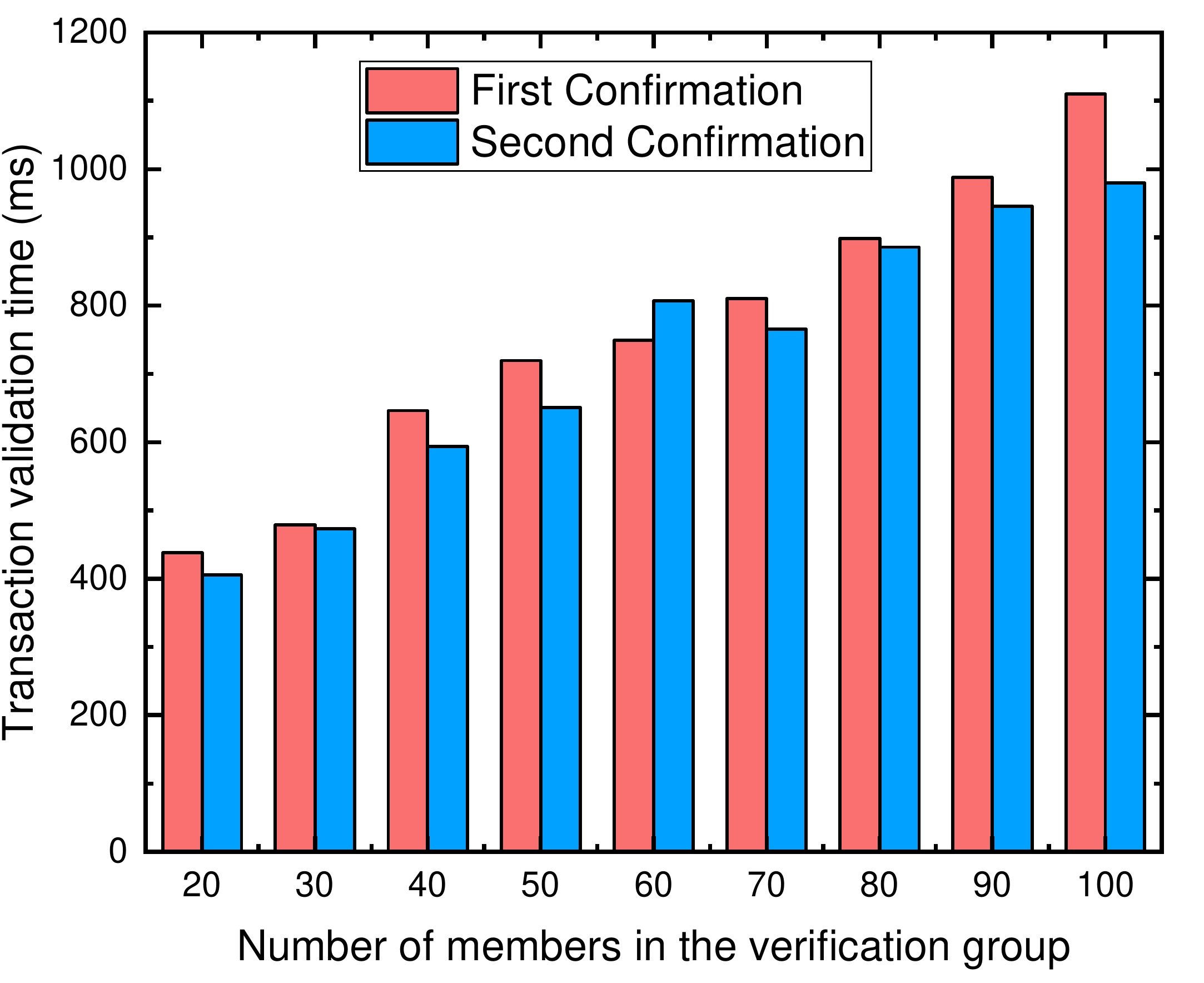}
     \caption{Time spent on the validation phase when the size of validation committee increases.}
    \label{figmembers}
\end{minipage}~~~
\begin{minipage}[t]{0.495\linewidth}
    \includegraphics[width=1.0\linewidth]{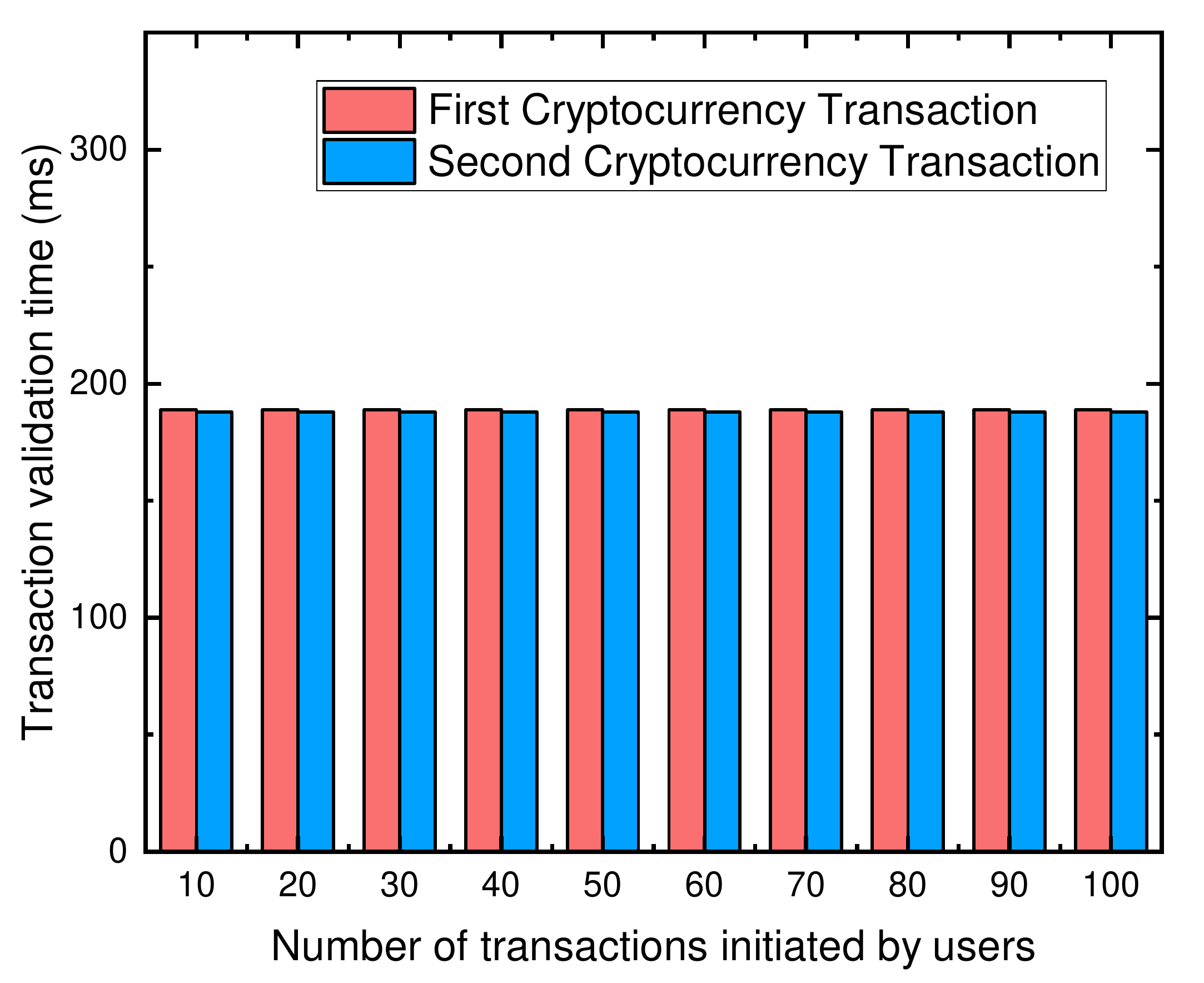}
    \caption{Time spent on the validation phase when the number of transactions between a pair of users changes.}
    \label{fig5}
\end{minipage}~~~
\end{figure*}

We also change the number of users in the validation committee and the number of transactions between a pair of users, so as to analyze the performance of the contract in the validation process. When we change the number of participants in the validation committee, the time which the contract takes to perform this step through \textit{TC.Validation} function is shown in Figure \ref{figmembers}. The first validation is for the transaction between payers and the intermediary $\mathcal{C}_1$, while the second validation is for the transaction between the intermediary $\mathcal{C}_2$ and payees. As we can see, when the number of validation committee members increases, the time required for the validation process also increases significantly. Nevertheless, even with 100 members in the validation committee, each validation process takes only about 1 second. This timing is negligible compared to the time it takes for transactions of public blockchain cryptocurrency like bitcoin to be confirmed. We can also see from Figure \ref{fig5} that as the number of transactions between a payer and a single payee increases, it will not increase the time cost of transaction verification through the contract. This is because all transactions between the same single user will be merged into one transaction through the contract, thereby not increasing the time cost.

\begin{figure*}
\centering
\begin{minipage}[t]{0.495\linewidth}
    \includegraphics[width=1.0\linewidth]{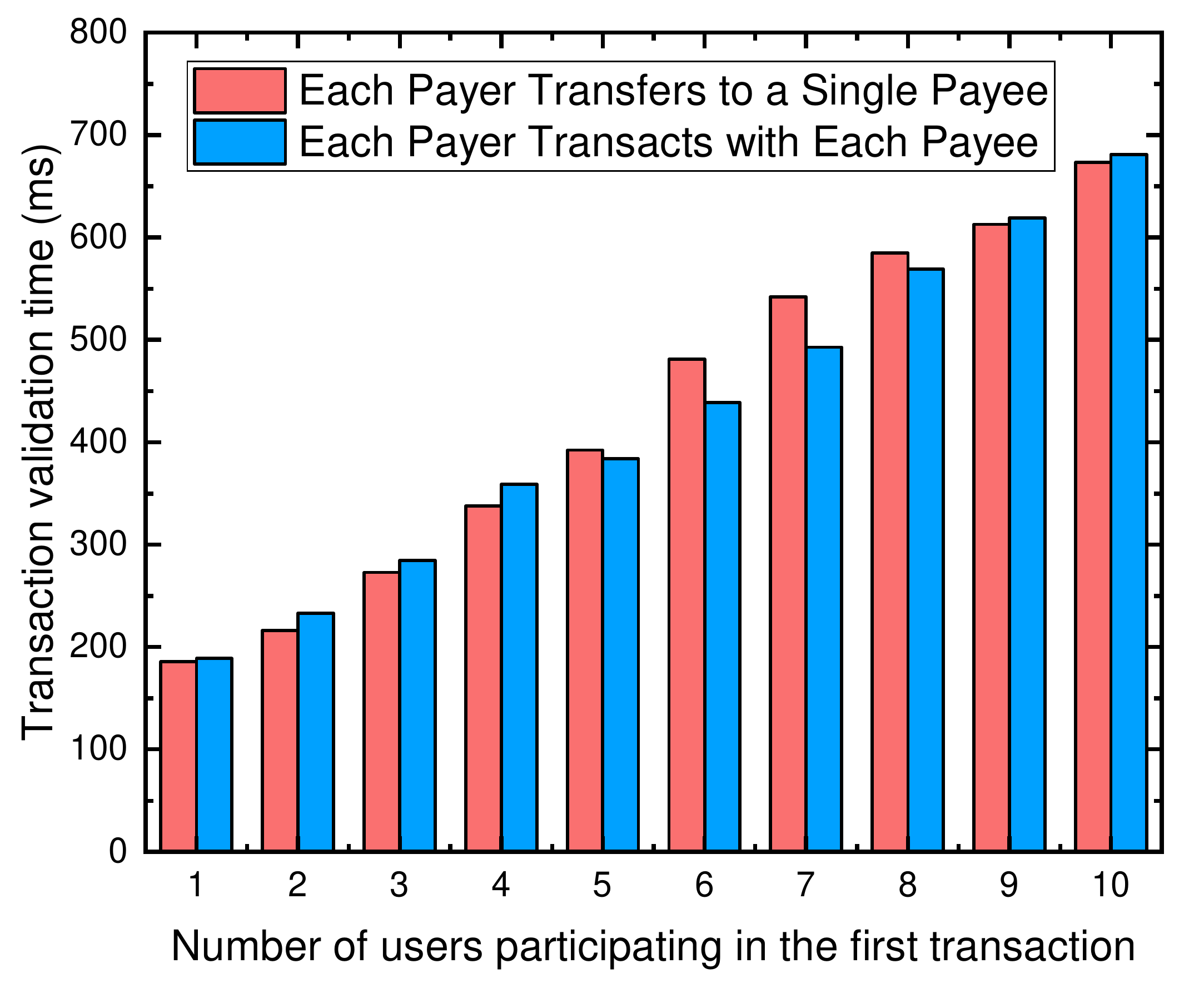}
\end{minipage}~~~
\begin{minipage}[t]{0.495\linewidth}
    \includegraphics[width=1.0\linewidth]{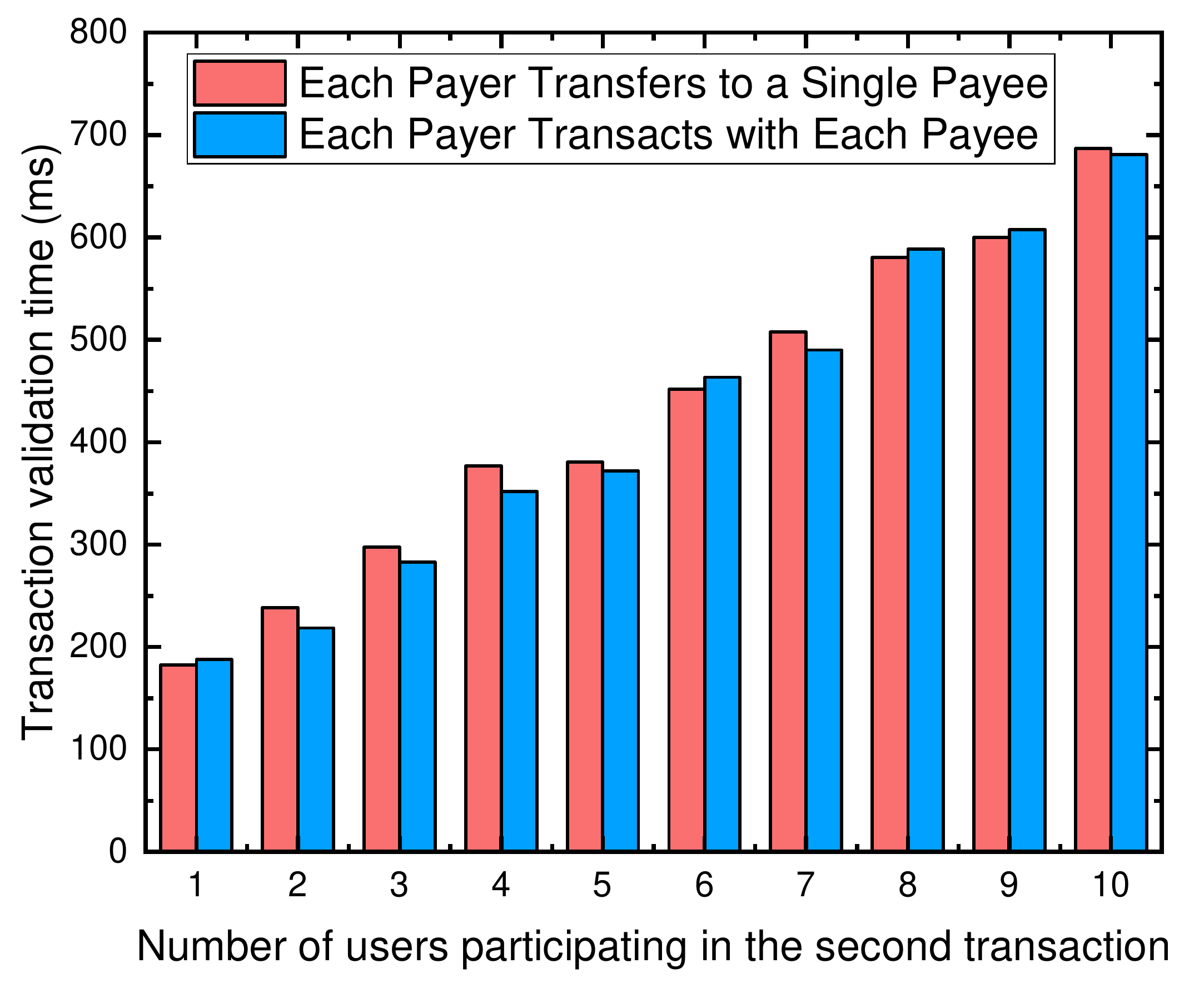}
\end{minipage}
\caption{Time spent on the validation phase of two cryptocurrency transactions.}
\label{fig6}
\end{figure*}

When we increase the number of users involved in the transaction, each additional transaction for each additional user adds to the runtime cost of the contract. For the convenience of the experiment, we add a pair of payer and payee at the same time. Figure \ref{fig6} shows the validation time required in two transactions respectively. The two colored columns in both figures represent two different cases. The first is the case where every payer transacts with a single payee, in which the number of transactions equals the number of participants. While the second case is when each payer transacts with each payee, in which the number of transactions is twice as large as the number of participants. However, in the range of allowable error, we can conclude from the figure that the time cost is the same in both cases. This means that the validation time cost of our scheme is only related to the number of users involved in the transaction, regardless of the number of transactions per user. This is because in our contract, the transactions involved in each participating user are merged, which is also the premise of our solution to maintain high throughput in the presence of a large number of transactions.

\section{CONCLUSION}
\label{conclusion}
In this paper, we proposed a decentralized cross-cryptocurrency exchange scheme between multiple users based on smart contracts. In our scheme, we use ether as a transit to link transactions between different kinds of two cryptocurrencies. We also implemented the contract and evaluated its execution overhead on our local machine. The results showed that the time cost of our scheme is only related to the total number of users involved in the transaction while the number of user's transactions has no significant impact on the runtime of our contract. In the future, we will improve our scheme from the experimental part in two aspects. Firstly, we will try to complete the implementation of the scheme with more users participation. Secondly, we will deploy our scheme on the Ethereum Mainnet to test the cost of our scheme under the actual network.

\bibliographystyle{unsrt}
\bibliography{Reference}
\end{document}